\documentclass[12pt, a4paper]{article}
\usepackage{amsmath}
\usepackage{amsthm}
\usepackage{amsfonts}
\usepackage{multicol}
\usepackage[dvips]{graphicx}
\usepackage{fixltx2e}

\newcommand{\danger}[1]{\textbf{#1}}

\input{epsf}             

\newtheorem{theorem}{Theorem}

\begin{document}

\title{\danger{Quantum Geometry I : \\
\normalsize{Basics of Loop Quantum Gravity \\} 
\small{The Quantum Polyhedra}}}
\author{\centerline{\danger{J. Manuel Garc\'\i a-Islas \footnote{
e-mail: jmgislas@iimas.unam.mx}}}  \\
Departamento de F\'\i sica Matem\'atica \\
Instituto de Investigaciones en Matem\'aticas Aplicadas y en Sistemas \\ 
Universidad Nacional Aut\'onoma de M\'exico, UNAM \\
A. Postal 20-726, 01000, M\'exico DF, M\'exico\\}

\maketitle

\begin{abstract}
General Relativity describes gravity in geometrical terms. This suggests that quantizing such theory is the same
as quantizing geometry. The subject can therefore be called quantum geometry
and one may think that mathematicians are responsible of this subject. Unfortunately, most
mathematicians are not aware of this beautiful area of study. 
Here we give a basic introduction to what quantum geometry 
means to a community working 
in a theory known as loop quantum gravity. It is directed towards
graduate or upper students of physics and mathematics.  We do it from a point of view of
a mathematician. 
\end{abstract}

\bigskip

\bigskip



\bigskip 

\bigskip

\bigskip

\bigskip 

\bigskip

\bigskip

\bigskip

\newpage

\section{Introduction}

The two most important physical theories of the 20th century are, of course,
general relativity and quantum mechanics. It is very well known to a physicist that
classical physical theories have a quantum version. The question then is: what is the
quantum version of general relativity? 
The thing is that there is no known correct answer to this question. 

Physically the question is what is quantum gravity? It is impressive that
such problem has been studied using many different directions; each one of these directions claims 
their theory is the solution to such problem.

Loop quantum gravity \cite{r}, \cite{c}, \cite{t} \cite{gp}, \cite{rv}
is one of these many directions. Among all of these directions, 
loop quantum gravity is the second most studied one, just after  
string theory.

There is no easy way to start learning loop quantum gravity. It is a difficult
theory, there is plenty of literature
out there most of which is very technical, and in fact there are many different
problems on which people are working.  

In this paper we give a basic introduction to only one of the constructions
of loop quantum gravity. We selected this particular problem because
we personally think it is the easiest one of all and, in fact, it is very beautiful. 

The problem is the following. It is known that general relativity is a theory of gravity 
which is described in geometrical terms. 
Therefore, quantizing general relativity must be equivalent to quantizing classical geometry.  

We can now rephrase the question: what is quantum geometry? And this is again a question with
no good answer, because 
quantizing classical geometry may mean a different
thing to different scientific communities. For example, it may mean something 
to a mathematician which is very different from what a physicist thinks.  
Mathematically speaking
quantum geometry may refer to a theory known as noncommutative geometry \cite{ac} or in fact, it may refer
to loop quantum gravity. There have been some studies relating some ideas of loop quantum gravity 
to noncommutative geometry. We do not explore this latter problem here.

The problem we introduce here is\footnote{The one we consider is the simplest one in order to start 
understanding the idea behind loop quantum gravity.}: if we want to start understanding what quantum
geometry may be, we should ask ourselves; is there a quantum version of a classical polyhedron. 
     
It turns out that the answer is yes. Classical polyhedra such as the Platonic solids for example
have a quantum version. This is very exciting in fact. However, unfortunately mathematicians are not
aware of this fact and it may be because this idea emerged in loop quantum gravity which is a theory
mostly invented by physicists. This is why our intention is to spread the idea to the mathematical community, 
and therefore
give a basic introduction from the point of view of a mathematician.

This review is directed from a mathematician point of view towards advanced undergraduates 
or postgraduates in mathematics and 
in physics. The mathematics of section 2 will be more familiar to mathematicians, whereas 
the mathematics of section 3 will be more familiar to physicists.      

We will start by recalling what classical polyhedra are and then will describe the quantum analogues.

\section{Classical Polyhedra}

This section is based on reference $\cite{a}$; however it is all written 
as we understand it, that is, our own words.
A classical polyhedron $P$ is just a solid in three dimensional space $\mathbb{R}^3$, ($P \subset \mathbb{R}^3$)
such that $\partial P$ 
is composed of  a finite number $k$
of flat polygons.\footnote{$\partial P$ denotes the boundary of the polyhedron $P$.} 
The polygons forming the polyhedron are called faces, and the sides and 
vertices of the faces are called edges and vertices. 
We denote the set of $k$ faces of the polyhedron by $f_1 , f_2, ...,f_k$.

We will restrict ourselves to convex polyhedra. Through each flat face $f_i$ of a classical polyhedron
there exists a plane $P_i$ that contains it. 
A convex polyhedron $\Pi$ is a classical polyhedron such that any two
polygonal faces $f_i \neq f_j$ are connected through other faces with common edges, and 
given a plane $P_i$ which contains the $f_i$ face, we have that $P_i \cap \Pi = f_i$ for all $i=1,...,k$.

We will consider bounded convex polyhedra, that is, polyhedra with bounded faces.

Given a convex polyhedron $\Pi$, consider $P_i$, the plane which contains
the face $f_i$. The unit vector $\mathbf{n}_i$ perpendicular to $P_i$ and pointing to the side
which does not contain any points of $\Pi$ is called the outward normal of $P_i$ relative to $\Pi$.

Now the most important theorem of this section.\footnote{In this theorem we use loop quantum 
gravity notation when referring to face areas.} 

\begin{theorem}[Minkowski] Let $\mathbf{n}_1, \mathbf{n}_2, ...., \mathbf{n}_k$ be unit vectors, $k \geq 4$,
such that any three different vectors $\mathbf{n}_i, \mathbf{n}_j, \mathbf{n}_{\ell} ,$ are linearly independent.  

Let  $A(f_1) , A(f_2),...., A(f_k) \in \mathbb{R}_{>0}$ such that 

\begin{equation}
\sum_{i=1}^k A(f_i) \mathbf{n}_i = 0 \nonumber
\end{equation}
Then there exists a closed convex polyhedron $\Pi$ with faces $f_i$ having areas 
$A(f_i)$ and outward normals $\mathbf{n}_i$.
\end{theorem}
This theorem implies that under given conditions there exist a convex polyhedron which satisfies the 
prescribed conditions.\footnote{For a proof of theorem 1 we refer the reader
to \cite{a}.}

However, it also turns out that given a convex
polyhedron whose faces $f_i$ have areas $A(f_i)$ and whose outward unit normals to the faces 
are $\mathbf{n}_i$,
the equation 

\begin{equation}
\sum_{i=1}^k A(f_i) \mathbf{n}_i = 0 
\end{equation}  
is satisfied. This is easy to see.  

Let $\mathbf{n}$ be a unit vector in $\mathbb{R}^3$. Consider the Euclidean inner product 
$< \mathbf{n} \ ,\mathbf{n}_i>$ of the unit vector $\mathbf{n}$ with all of the unit normals to the faces 
of the convex polyhedron $\Pi$. The number $< \mathbf{n} \ ,\mathbf{n}_i>A(f_i)$ is the area of the projection of 
the face $f_i$ into a plane $\mathbf{n}^{\perp}$ whose all vectors are orthogonal to $\mathbf{n}$. All of the unit vectors 
$\mathbf{n} , \mathbf{n}_1 , ....,\mathbf{n}_k \in S^2$, where 

\begin{equation}
S^2 = \{ (z , x, y) \in \mathbb{R}^3 \mid z^2 + x^2 + y^2 =1 \} \nonumber
\end{equation}  
is the unit sphere; therefore we have that some of the interior products $< \mathbf{n} \ ,\mathbf{n}_i>$ will be positive, 
and some others will be negative, since some of the $\mathbf{n}_i$ point in the same direction as $\mathbf{n}$
and some point in the opposite direction. The projection to the plane $\mathbf{n}^{\perp}$
of the faces whose normal vectors $\mathbf{n}_i$ 
point in the same direction as $\mathbf{n}$, and the projection of those 
whose normal vectors $\mathbf{n}_i$ point in the opposite direction as $\mathbf{n}$ cover the same area.
Then 

\begin{equation}
\sum_{i=1}^k < \mathbf{n} \ ,\mathbf{n}_i> A(f_i) = 0 \nonumber
\end{equation}  
$\Rightarrow$

\begin{equation}
< \mathbf{n} \ , \sum_{i=1}^k A(f_i) \mathbf{n}_i > = 0 \nonumber
\end{equation} 
Since $\mathbf{n} \neq \vec{0}$, and it is an arbitrary vector, we have that 

\begin{equation}
\sum_{i=1}^k A(f_i) \mathbf{n}_i = 0 \nonumber
\end{equation}  
Therefore equation $(1)$ is proved.

\section{Quantum Polyhedra}

Classical physical theories have a quantum version.\footnote{For instance, a quantum version
of space exists. See for example \cite{bhh}.}
The question is: can mathematics be quantized? 
Well, let us start by asking, is there a quantum version of a classical convex polyhedron described in the previous 
section?

Surprisingly there is, and loop quantum gravity has described these quantum versions 
\cite{rv}, \cite{bds} \cite{js}.
But the idea can be generalised to quantizing any convex polyhedron. In this section we describe 
the quantum version of a classical convex polyhedron. We give a very basic introduction
to this beautiful subject. It is in fact a hard thing to do since the subject is full of
difficult and advanced mathematics. At least in this first introduction paper we keep it simple.  
This section is mathematically inspired in reference \cite{bh}.

There is an action of the Lie group $SO(3)$ of rotations in the Euclidean space $\mathbb{R}^3$, and the
areas $A(f_i)$ of the faces $f_i$ of the convex polyhedron $\Pi$ as well as its volume remain invariant
under such rotations. The set of the normal unit vectors $\mathbf{n}_i$ to the faces obviously remain unit vectors.  

Quantizing a convex polyhedron is defined by assigning a Hilbert space $\mathcal{H}_i$ to each of its faces $f_i$
and the tensor product $\mathcal{H}_1 \otimes \mathcal{H}_2 \otimes \cdot \cdot \cdot \cdot \otimes \mathcal{H}_k$ 
to the polyhedron in the following way. This implies that the observables are related to measures on the
quantum polyhedron faces. In classical geometry a polyhedron $\Pi$ has faces $f_i$ of certain area
$A(f_i)$. Area in classical geometry is a classical observable. Therefore its quantum counterpart
is called quantum area and it must be an operator defined on a Hilbert
space. This is understood as follows. 

As $SO(3)$ sends the unit sphere $S^2$ to itself, the Hilbert space associated
to each face is in fact $\mathbf{L}^2(S^2)$, the space of complex valued 
squared-integrable functions.\footnote{$SO(3)$ not only acts on the Hilbert space $\mathbf{L}^2(S^2)$,
it can also act on $\mathbf{L}^2(\mathbb{R}^3)$ for instance. However we choice the
action restricted to $\mathbf{L}^2(S^2)$ since our equations will not depend on
the radial coordinate. } That is,

\begin{equation}
\mathbf{L}^2(S^2) = \bigg\{ \psi : S^2 \rightarrow \mathbb{C} \mid \int_{S^2} 
\mid \psi(\vec{x}) \mid^2 d\vec{x}< \infty \bigg\} \nonumber
\end{equation}  
Just as the Lie group of rotations $SO(3)$ acts in $S^2$, it also acts in the Hilbert space 
$\mathbf{L}^2(S^2)$ by 

\begin{equation}
R \ \psi(\vec{x}) := \psi(R^{-1} \ \vec{x}) \nonumber
\end{equation}  
where $R : S^2 \rightarrow S^2$ is a rotation and $R^{-1}$ is its inverse.

To the polyhedron we assign the tensor product $\otimes_{k} \ \mathbf{L}^2(S^2)$, so that a vector
in $\otimes_{k} \ \mathbf{L}^2(S^2)$ is given by 
$\psi_1(\vec{x_1}) \otimes \psi_2(\vec{x_2})  \otimes \cdot \cdot \cdot \cdot \otimes \psi_k(\vec{x_k})$
such that the $SO(3)$ action on this tensor product space is given by

\begin{equation}
R \  (\psi_1(\vec{x_1}) \otimes \psi_2(\vec{x_2})  \otimes \cdot \cdot \cdot  \otimes \psi_k(\vec{x_k}))
:= \psi_1(R^{-1} \ \vec{x_1}) \otimes \psi_2(R^{-1} \ \vec{x_2})  \otimes \cdot \cdot \cdot \otimes 
\psi_k(R^{-1} \ \vec{x_k}) 
\nonumber
\end{equation}  
Physically, the wave function of a quantum polyhedron is a complex valued function
defined on the tensor product $\otimes_{k} \ \mathbf{L}^2(S^2)$ Hilbert space, such that
$\psi_1(\vec{x_1}) \otimes \psi_2(\vec{x_2})  \otimes \cdot \cdot \cdot \cdot \otimes \psi_k(\vec{x_k})$
is a unit vector in the Hilbert space $\otimes_{k} \ \mathbf{L}^2(S^2)$.
Mathematically this is written

\begin{equation}
\int_{S^2 \times S^2 \cdot \cdot \cdot \times S^2} \mid
\psi_1(\vec{x_1}) \otimes \psi_2(\vec{x_2})  \otimes \cdot \cdot \cdot \cdot \otimes \psi_k(\vec{x_k})
\mid^2 d\vec{x_1} \ d\vec{x_2} ....d\vec{x_k} = 1\nonumber
\end{equation}
where the integral is over $k$ products of $S^2$.
In fact this latter integral is given by
$\prod_{i=1}^k \int_{S^2} \mid
\psi_i(\vec{x_i}) \mid^2  d\vec{x_i}$.

\bigskip

When we study quantum mechanics, we know that the wave function of a system is a superposition
(linear combination of basis vectors in the Hilbert vector space) of states which are eigenvectors 
of an observable (self-adjoint operator in the Hilbert space). The eigenvalues of the observable 
are the observed quantities with a certain probability.

When we say that the quantum polyhedron has a wave function, or is in the state
$\psi_1(\vec{x_1}) \otimes \psi_2(\vec{x_2})  \otimes \cdot \cdot \cdot \cdot \otimes \psi_k(\vec{x_k})$,
we must understand that it is a superposition state. What are the observed quantities? What is
an observable in this theory of quantum polyhedra? 
In order to answer these questions we should know some more things. Let us discuss these issues.

The Hilbert space $\mathbf{L}^2(S^2)$ of squared-integrable functions over $S^2$ has an inner
product given by

\begin{equation}
< \psi(\vec{x}) \mid \chi (\vec{x}) > \ = \int_{S^2} \overline{\psi(\vec{x})} \ \chi(\vec{x}) \ d\vec{x} \nonumber
\end{equation}
If we introduce spherical coordinates in $S^2$ 

\begin{equation}
f(\theta , \phi) = (\cos \theta  , \ \sin \theta \cos \phi  , \ \sin \theta \sin \phi) \nonumber
\end{equation}
such that $0 < \theta < \pi , 0 < \phi < 2 \pi$.
Then the functions $\psi(\vec{x})$ become functions
of the spherical angles $\psi(\theta , \phi) $ and
the inner product can be written explicitly as

\begin{equation}
< \psi(\theta , \phi) \mid \chi (\theta , \phi) > \ = \frac{1}{4 \pi}
\int_{S^2} \overline{\psi(\theta , \phi)} \ \chi(\theta , \phi)  \sin \theta \ d\theta \ d\phi \nonumber
\end{equation}
$4\pi$ is the area of the unit sphere, or in other words.\footnote{\ \ \ $\frac{1}{4 \pi}
\int_{S^2} \sin \theta \ d\theta \ d\phi = 1$}

In this Hilbert space the observables include the self-adjoint operators 
$J_1 , J_2, J_3 : \text{Dom}(\mathbf{L}^2(S^2)) \rightarrow \mathbf{L}^2(S^2)$ given by

\begin{equation}
J_1 = i \ \bigg( \sin \phi \ \frac{\partial}{\partial \theta} 
+ \cos \phi \ \frac{\cos \theta}{\sin \theta} \ \frac{\partial}{\partial \phi} \bigg) \nonumber
\end{equation}

\begin{equation}
J_2 = i \ \bigg( - \cos \phi \ \frac{\partial}{\partial \theta} 
+ \sin \phi \ \frac{\cos \theta}{\sin \theta} \ \frac{\partial}{\partial \phi} \bigg) \nonumber
\end{equation}

\begin{equation}
J_3 = - i \ \frac{\partial}{\partial \phi}  \nonumber
\end{equation}
and the commutation relations of these operators are given by

\begin{equation}
[ J_1 , J_2 ] = i \ J_3 \ \ \ , \ \ \ [ J_2 , J_3 ] = i \ J_1 \ \ \ , \ \ \ [ J_3 , J_1 ] = i \ J_2 \nonumber
\end{equation}
There is also an operator known as the Casimir operator given by  

\begin{equation}
J^2 = J_1^2 + J_2^2 + J_3^2 \nonumber
\end{equation}
Using the expressions for $J_1 , J_2 , J_3$ it can be seen that 

\begin{equation}
J^2 = - \frac{1}{\sin \theta} \ \bigg( \frac{\partial}{\partial \theta} 
\bigg( \sin \theta \frac{\partial}{\partial \theta} \bigg) + \frac{1}{\sin \theta} 
 \frac{\partial^2}{\partial \phi^2} \bigg) \nonumber
\end{equation}
This latter expression is the minus Laplacian on the sphere which has eigenvectors given by the
well known spherical harmonics functions $Y(\theta , \phi)$.
This means that

\begin{equation}
J^2 \ Y(\theta , \phi) = j (j+1) \ Y(\theta , \phi) 
\end{equation}
where $j \in \mathbb{Z}_{\geq 0}$. Each eigenvalue $j (j+1)$ is 
of multiplicity $2j +1$ and therefore the eigenvectors
of the operator $J^2$ with eigenvalue $j (j+1)$ generate a subspace $H_j$ of $\mathbf{L}^2(S^2)$.
This implies that the Hilbert space $\mathbf{L}^2(S^2)$ is a direct sum given by

\begin{equation}
\mathbf{L}^2(S^2) = \bigoplus_{j=0}^{\infty} \ H_j \nonumber
\end{equation}
It is customary to denote the orthogonal basis of eigenvectors with eigenvalue $j (j+1)$ that generate the
subspace $H_j$ by $Y_{m}^{j}( \theta , \phi)$ where $m$ takes integer values $- j \leq m \leq j$. 

In loop quantum gravity
the observable $J$ is the area operator,\footnote{The relation of the observable operator $J$
and an area operator is a construction derived in loop quantum gravity.  
This relation derivation is out of the scope of this review and we do not plan
to deal with it at the moment. However it is our intention to have a new review in a future and it will be 
explained there.} 
and formula $(2)$ is interpreted physically as the squared
area of face $f_i$ of the quantum polyhedron $\Pi$. 
Face $f_i$ has therefore quantized area given by the numbers

\begin{equation}
A(f_i) = \sqrt{j_i (j_i +1)} \nonumber
\end{equation}

On the other hand, a general vector $\psi(\theta , \phi)$ (wave function) in the Hilbert space $\mathbf{L}^2(S^2)$ is
a linear combination of bases vectors (superposition) given by

\begin{equation}
\psi(\theta , \phi) = \sum_{j=0}^{\infty} \sum_{m= -j}^{j}  c_{m}^{j} \ Y_{m}^{j}( \theta , \phi) \nonumber
\end{equation}
where $c_{m}^{j} \in \mathbb{C}$.

A wave function of a quantum polyhedron is given by

\begin{equation}
\psi_1(\theta_1 , \phi_1) \otimes \psi_2(\theta_2 , \phi_2) \otimes \cdot \cdot \cdot \otimes \psi_k(\theta_k , \phi_k)  \nonumber
\end{equation}
where $k$ is the number of faces of the classical polyhedron. It is of course a
linear combination(superposition of states) of basis vectors 
which can be written as

\begin{eqnarray}
\bigotimes_{i=1}^{k} \psi_i(\theta_i , \phi_i) = \sum_{j_i=0}^{\infty} \sum_{m_i= -j_i}^{j_i}  
\prod_{i=1}^{k} c_{m_i}^{j_i} \ \bigotimes_{i=1}^{k}
 Y_{m_i}^{j_i}( \theta_i , \phi_i)  \nonumber
\end{eqnarray}

After a measurement of the observable $J$ the quantum polyhedron will be in a particular state 

\begin{equation}
Y_{m_1}^{j_1}( \theta_1 , \phi_1) \otimes Y_{m_2}^{j_2}( \theta_2 , \phi_2) 
\otimes \cdot \cdot \cdot \otimes Y_{m_k}^{j_k}( \theta_k , \phi_k) \nonumber
\end{equation}
This implies that we have a quantum polyhedron which area faces are quantized and the total area
of the quantum surface is\footnote{We have studied a very simplified
problem. We have not dealt for instance with more complicated mathematics
behind quantum polyhedra, like the theory of representations, including the quantum version of classical formula $(1)$.
We will deal with this in a future review.} 

\begin{equation}
A(\Pi) = \ell_{P}^2 \sum_{i =1}^{k} \sqrt{j_i (j_i +1)} \nonumber
\end{equation}
where $\ell_{P}$ is the Planck length and it is introduced in the previous formula in order to have the correct dimensions.

\section{Conclusions}

This short review was intended to be a simple first introduction to one
particular subject of loop quantum gravity; quantum polyhedra. It was
directed to undergraduate or to first year postgraduate students in physics and mathematics. 
It was our intention
to describe it from the perspective of a mathematician, and we hope we have succeeded in this task.

It is our intention to continue introducing loop quantum gravity to mathematicians, since
most mathematicians are not aware of the beautiful subject called loop quantum gravity.

As this is a first introduction we have left so many things out; loop quantum gravity is a very extensive field
and no first introduction will be satisfactory. Even dealing with quantum polyhedra
requires more formal, and advanced mathematics we have not dealt with. 

From what we studied in this first introduction, we have learnt that quantum polyhedra 
states are superposed and once we have performed a measure of its faces areas 
the superposition collapses to a polyhedron which faces have discrete areas. 
This means that the area operator is quantised and therefore we have a first glimpse 
of what quantum geometry is form the perspective of loop quantum gravity. 

When quantizing geometry, 
area is not continuous but discrete. It happens the same when considering a volume
operator and finding that its spectrum is discrete. We did not consider the volume operator
here, since it is more complicated. But physicists of loop quantum gravity interpret
the discrete spectrums as thinking of space formed by quantum entities called quanta of space.

\end{document}